\documentclass[aps,prl,twocolumn,superscriptaddress]{revtex4}
\usepackage{graphicx}
\usepackage{color}

\begin{document}


\title{Optical transition of the $^{229}$Th nucleus in a solid-state environment}


\author{Wade~G.~Rellergert}
\affiliation{Department of Physics and Astronomy, University of California, Los Angeles, California 90095, USA}
\author{D.~DeMille}
\affiliation{Department of Physics, Yale University, New Haven, Connecticut 06511, USA}
\author{R.~R.~Greco}
\affiliation{Los Alamos National Laboratory, Los Alamos, New Mexico 87545, USA}
\author{M.~P.~Hehlen}
\affiliation{Los Alamos National Laboratory, Los Alamos, New Mexico 87545, USA}
\author{J.~R.~Torgerson}
\affiliation{Los Alamos National Laboratory, Los Alamos, New Mexico 87545, USA}
\author{Eric~R.~Hudson}
\affiliation{Department of Physics and Astronomy, University of California, Los Angeles, California 90095, USA}
\email[]{eric.hudson@ucla.edu}


\date{\today}

\begin{abstract}
We describe a novel approach to directly measure the energy of the narrow, low-lying isomeric state in $^{229}$Th. Since nuclear transitions are far less sensitive to environmental conditions than atomic transitions, we argue that the $^{229}$Th optical nuclear transition may be driven inside a host crystal with a high transition Q.  This technique might also allow for the construction of a solid-state optical frequency reference that surpasses the precision of current optical clocks, as well as improved limits on the variability of fundamental constants.  Based on analysis of the crystal lattice environment, we argue that a precision of $3\times10^{-17}<\Delta f/f <1\times10^{-15}$ after 1~s of photon collection may be achieved with a systematic-limited accuracy of $\Delta f/f \sim 2 \times 10^{-16}$. Improvement by $10^2-10^3$ of the constraints on the variability of several important fundamental constants also appears possible.
\end{abstract}


\maketitle


The enormous effort towards building next-generation atomic clocks based on optical transitions~\cite{takamoto_spectroscopy_2003, ido_precision_2005, barber_direct_2006, rosenband_observation_2007} testifies to the impact a more accurate time reference will have. Already, atomic clock experiments have provided some of the most stringent tests of General Relativity~\cite{ashby_testing_2007} and produced the tightest constraints on present day variation of many of the fundamental constants~\cite{rosenband_frequency_2008}.
Despite their successes, traditional clock experiments are cumbersome: the high-quality (Q) oscillator is an atomic transition, which is  sensitive to its environment. To mitigate environmental influences, modern clock experimenters routinely employ complicated interrogation schemes, \textit{e.g.} atomic fountains or 3-dimensional optical lattice confinement. An interesting shift in paradigm is to consider an optical clock based on a nuclear transition. Just as in atomic clocks, the high-Q oscillator, \textit{i.e.} the nuclear transition, can be addressed by laser spectroscopic techniques -- as long as the transition energy is accessible with current laser technology -- with the added benefit, due to screening by the atomic electrons, of reduced sensitivity to the environment. The low-lying isomeric state in $^{229}$Th provides such a transition.

The existence of a low-lying state in $^{229}$Th, with an estimated natural linewidth, $\Gamma_n$, in the range $0.1 \lesssim \Gamma_n/2 \pi \lesssim  10$~mHz~\cite{tkalya_decay_2000, ruchowska_nuclear_2006}, was indirectly established over 3 decades ago through measurements of $\gamma$-ray energies resulting from $\alpha$-decay of $^{233}$U~\cite{kroger_features_1976}.  Subsequent indirect measurements  determined the isomeric state ($\rm{I} = 3/2^+$) to be 3.5~eV above the ground state ($\rm{I} = 5/2^+$)~\cite{helmer_excited_1994};
however, direct searches for UV emission from this M1 transition were unsuccessful~\cite{irwin_observation_1997, richardson_ultraviolet_1998}. Recently, a new indirect measurement has established that the transition energy is instead 7.6~$\pm$~0.5~eV~\cite{beck_energy_2007}.  Although well into the vacuum ultraviolet (VUV), this energy is amenable to study by laser spectroscopy.

Here, we propose a novel approach to \textit{directly} measure the transition energy to higher precision by interrogating $^{229}$Th that has been doped into a VUV transparent crystal~\cite{hudson_investigation_2008, peik_prospects_2009}. This approach has several advantages.  First, in contrast to current proposals~\cite{peik_nuclear_2003, steele_laser_2008} using laser-cooled ions that can address densities of at most 10$^8$ $^{229}$Th nuclei/cm$^3$, this method results in samples containing densities as high as 10$^{19}$ $^{229}$Th nuclei/cm$^3$ ($\leq$100 $\mu$Ci/cm$^3$). The ability to address such large numbers of absorbers leads to fluorescence rates sufficient for the use of a synchrotron light source to directly measure the transition energy. Given that the transition energy is only known with $\sim$25 nm accuracy and is in the VUV, this is a dramatic experimental simplification. Second, as is known from M\"{o}ssbauer spectroscopy, nuclear transitions are far less sensitive to their surroundings than atomic transitions.  Thus a sample of $^{229}$Th atoms doped into a suitable host crystal retains a narrow transition linewidth, potentially allowing for the construction of a solid-state optical frequency reference with a higher precision than current optical clocks.  Third, $^{229}$Th atoms in a crystal lattice are confined to the Lamb-Dicke regime~\cite{dicke_effect_1953} - that is, the recoil energy is much less than the energy required to create a phonon in the lattice, and therefore the internal and external degrees of freedom are decoupled.  As a result, the transition energy will not be sensitive to recoil or first order Doppler effects. Nonetheless, there are still a variety of small effects that must be considered.  In what follows, we will show that as long as each nucleus experiences the same crystal lattice environment, many of these effects lead to constant, identical shifts rather than broadenings of the transition frequency.  As a result, they should not affect the performance of a clock based on this transition.


In the absence of external fields, the Hamiltonian for a $^{229}$Th nucleus in a crystal can be written as $\rm H = H_{0} + H_{FS} + H_{HFS}$, where H$_{0}$ is the single-particle nuclear Hamiltonian, \textit{e.g.} the Nilsson Hamiltonian~\cite{casten_nuclear_1990}, while H$_{\rm{FS}}$ and H$_{\rm{HFS}}$ are the Hamiltonian describing the fine and hyperfine interactions, respectively, arising from the crystal environment. As will be discussed later, we are concerned only with ionically bonded crystals with no unpaired electron spins, therefore the fine-structure interaction vanishes. Thus, the hyperfine interactions are the dominant source of shifts and broadenings of the nuclear transition.  Using a multi-pole expansion, the hyperfine Hamiltonian can be written as $\rm H_{HFS} = H_{E0} + H_{M1} + H_{E2} + \cdots$ .  These are the same effects that arise in M\"{o}ssbauer and NMR spectroscopy, and as a result, each can be reliably estimated using previous measurements.  In what follows, we explain the estimated effect of the first few terms in the series for a generic lattice environment.

The electric monopole term, H$_{\rm{E0}}$, describes the contact interaction between the nucleus and its electron cloud.  Thus, changes in the size of either the electron cloud or the nucleus lead to changes in the transition energy. Since the electron density at a given lattice site is temperature dependent, the transition energy will shift with temperature, and any temperature gradients across a sample containing $^{229}$Th can lead to inhomogeneous line broadening. M\"{o}ssbauer spectroscopy studies indicate that this shift will be $\lesssim$100~MHz compared to free ions, and that the temperature dependence of the shift will be on the order of 10~kHz/K~\cite{gibb_principles_1980, schroeer_isomer_1979, rohde_temperature_1983}.  If the atoms reliably substitute for a single species in a crystal the shift will be the same for all $^{229}$Th nuclei.
Additionally, as laboratory techniques can lead to temperature stability better than 0.1~mK across the crystal~\cite{davidowitz_novel_1995}, this term contributes $\sim$1~Hz to the transition linewidth.

The magnetic dipole term, H$_{\rm{M1}}$, will give rise to a shift and splitting of the ground and excited state energies if there is a non-zero magnetic field at the nucleus.  Crystals that are transparent in the VUV, and thus suitable hosts for these studies, necessarily have large conduction band gaps and therefore will be free of spurious currents.  Furthermore, in ionically bonded crystals of interest, thorium substitutes as a Th$^{4+}$ ion (isoelectronic with radon), so all electron spins will be paired.  Therefore, the only magnetic fields in the crystal will be those due to the magnetic moments of other nuclei.  The magnetic field produced by one nuclear magneton at the shortest typical distance ($\sim$2.5~\AA) in a crystal is $\sim$400~mG.  As the magnetic dipole moment of the 5/2$^{+}$ and 3/2$^{+}$ states are estimated to be 0.45~$\mu_N$ and -0.08~$\mu_N$ respectively~\cite{dykhne_matrix_1997}, this would lead to a broadening of the transition energy by $\sim$400~Hz.  Given the contributions of all neighboring nuclei, this could lead to a broadening of 1-10~kHz and thus could be the dominant source of broadening.  However, it appears this broadening can be reduced by a factor of 20 with the M\"{o}ssbauer analog~\cite{anisimov_concept_2007} of  ``magic angle" spinning used in NMR.

The electric quadrupole term, H$_{\rm{E2}}$, can also give rise to a shift and splitting in the transition energy if an electric field gradient is present since both states have spin I~\textgreater~1/2 and thus support an electric quadrupole moment.  The electric field gradient at the nucleus depends sensitively on the structure of the host crystal and the effect of this term must be considered on a case by case basis.  Further complicating a reliable estimate of the effect is that only the ground state quadrupole moment of $^{229}$Th is known (4.3~barns~\cite{gerstenkorn_hyperfine_1974}).  Nevertheless, an estimate for the size of the effect can be obtained by considering only the ground state shift.  Field gradients in crystals are typically in the range of 10$^{16}$-10$^{21}$~V/m$^2$~\cite{ramakrishna_electric_1966, roesch_electric_1975}, leading to transition energy shifts of 1~kHz-100~MHz. However, as long as the $^{229}$Th nuclei reliably substitute into a single lattice site, this interaction results in a constant shift of the transition frequency  -- with a temperature ($\rm{T}$) dependence of $\Delta f/f~=~\alpha\Delta \rm{T}$ where $\alpha$ is the crystal coefficient of linear expansion. We note that H$_{\rm{E2}}$ is minimized for host crystals with cubic or icosahedral symmetry~\cite{knop_arrangements_1975}.

\begin{table}
\begin{ruledtabular}
\begin{tabular}{l c c}
  Effect & T-dependent Shift & Broadening\\
  \hline
  Monopole & $5\times10^{-12}$K$^{-1}$ & $5\times10^{-12}$K$^{-1}$ \\
  Magnetic Dipole & 0 & $\leq$$5\times10^{-12}$ \\
  Electric Quadrupole & $\leq$$1\times10^{-12}$K$^{-1}$ & $\leq$$1\times10^{-12}$K$^{-1}$ \\
  Second Order Doppler & $\leq$$1\times10^{-15}$K$^{-1}$ &  $\leq$$1\times10^{-15}$K$^{-1}$ \\
\end{tabular}
\end{ruledtabular}
\caption{\label{ShiftSummary} Summary of relevant fractional shifts and broadenings for LiCAF. For broadenings, K$^{-1}$ refers to the magnitude of the temperature variation across the sample.}
\end{table}

%

Because nuclei can vibrate in their lattice sites these interactions could lead to a broadening of the transition frequency if the zero-phonon transitions are mode dependent.  Such dependence is likely only for the optical phonon modes where the $^{229}$Th nuclei are vibrating relative to their nearest neighbors, however, even at room temperature these modes are expected be frozen out~\cite{slack_thermal_1961}. Further, such an effect would lead to resolvable sidebands if the shift is more than the linewidth.

Finally, the vibration of the nuclei also leads to second order Doppler effects which can cause a shift in the transition energy.  Using a simple harmonic oscillator model this effect can be estimated as $ \Delta f / f = 3 k T / (2 M c^2)$~\cite{gibb_principles_1980}, yielding a temperature dependent shift on the order of 1~Hz/K; $\Delta f \leq 0.1$~mHz across the sample for the assumed temperature stability and uniformity.

While the above analysis shows that the crystal lattice environment causes large, perhaps incalculable, shifts in the transition frequency, this modified transition frequency (in the absence of defects) is no less fundamental than the atomic frequencies currently used for time keeping purposes.  Thus, the only effects limiting the usefulness of this transition for a primary frequency standard are the intrinsic broadenings and environmental effects that shift the nuclear energies, summarized in Table~\ref{ShiftSummary} for LiCAF (see below). The former will impact the precision of the frequency measurement, while the latter impacts its accuracy. We show below that even in the worst case scenario (10~kHz broadening due to the magnetic dipole term), the large number of absorbers allows the frequency to be measured with a higher precision than current clocks. As standard laboratory techniques lead to suppression of electric and magnetic fields to below 1~mV/cm and 1~$\mu$G respectively, the most important environmental effect appears to be the crystal temperature; a 1~mK temperature difference results in a fractional shift of the frequency by 6$\times 10^{-15}$.  With state of the art thermometer accuracies of $\sim$0.04~mK~\cite{ITS90}, a clock based on this architecture may not possess the necessary accuracy to be a primary standard; however, it should display high short-term stability which would make it useful as a flywheel oscillator that could replace \textit{e.g.} the hydrogen maser standard.


In addition to its potential use as a frequency reference, the transition also displays enhanced sensitivity to fundamental constant variation compared to measurements based on electronic transitions~\cite{flambaum_enhanced_2007,he_enhanced_2007}.
The dependence of the transition frequency on the fine-structure constant ($\alpha$) and the ratio of the light and strange quark mass to the QCD length scale ($X_q = m_q/\Lambda_{QCD}$ and $X_s = m_s/\Lambda_{QCD}$, respectively), is derived from H$_0$ as~\cite{flambaum_enhanced_2006, he_enhanced_2007}:$~\Delta f~\approx~\left(c_{\alpha} \frac{\delta \alpha}{\alpha} + c_{q} \frac{\delta X_q}{X_q} + c_{s} \frac{\delta X_s}{X_s}\right)$
where $c_{i}$ is a model dependent constant.  Using the constants derived in Ref.~\cite{he_enhanced_2007}, a variation in $\alpha$, $X_q$, or $X_s$ of 1 part in 10$^{17}$ will lead to a change of the transition frequency by $\sim$150~Hz, $\sim$3~kHz, and $\sim$37~kHz, respectively. 
Hence, measurements of the transition frequency at the 1 Hz level (see below) separated by one year could determine independent fractional variation of these fundamental constants at the $<$$10^{-19}$~yr$^{-1}$, $10^{-20}$~yr$^{-1}$, and $10^{-21}$~yr$^{-1}$ levels respectively; a 2-3 order of magnitude improvement over current limits~\cite{rosenband_frequency_2008,flambaum_enhanced_2009,flambaum_variation_2008}.


The first steps towards these exciting applications are identifying suitable host crystals and determining the transition frequency to a greater precision. The key requirements for the host crystal are that it be reasonably transparent in the VUV, have a pure crystalline structure, and chemically accept Th$^{4+}$ ions. Numerous candidate hosts can be identified in the extensive family of fluoride crystals.
Potential crystals include Na$_2$ThF$_6$, LiCAF (LiCaAlF$_6$), LiSAF (LiSrAlF$_6$), YLF (LiYF$_4$), and CaF$_2$, which can be grown by various techniques, \textit{e.g.} Czochralski, Bridgman, etc. ~\cite{Burkhalter2001, Samtleben2005}. While these crystals have band gaps of $>$10\,eV, impurities and color centers can cause residual absorption and may interfere with the $^{229}$Th optical spectroscopy. Various metal ions as well as oxygen-based impurities at the part-per-million level can lead to absorption coefficients of up to 0.1\,cm$^{-1}$~\cite{Sabatini1975}.  To ensure crystal transparency, standard purification methods can be employed to reduce these impurities to part-per-billion levels~\cite{Hehlen2009}. We are currently applying these methods to the preparation of high purity metal fluoride precursors for crystal growth.


Thus far our experimental effort has focused on LiCAF which is transparent down to 110~nm and has all electron spins paired. $^{229}$Th$^{4+}$ ions are expected to substitute at the Ca$^{2+}$ location in LiCAF along with two F$^-$ interstitials for charge compensation~\cite{jackson_private_2008}. Due to the octahedral symmetry of the Ca site, electric field gradient effects will be minimized. We have fabricated a Th:LiCAF test crystal using the less expensive $^{232}$Th isotope. A density in excess of 10$^{18}$ $^{232}$Th/cm$^3$ was measured using Rutherford back-scattering.  We have also made measurements of the fluorescence backgrounds which might arise from the $\alpha$-decay of $^{229}$Th (4.8\,MeV, $\tau$~=~7880\,yr) by bombarding this crystal with a 30~nA, 4.8~MeV beam of $\alpha$-particles at the Ion Beam Materials Laboratory of Los Alamos National Laboratory.  A total fluorescence rate corresponding to 0.3~photons per $\alpha$-particle was observed over the unfiltered PMT bandwidth (Hamamatsu R7639; 115 nm - 230 nm).  Thus, a crystal containing 10$^{18}$~$^{229}$Th nulcei yields a background fluorescence rate of 2~MHz.  Even in the unrealistic case that all of this fluorescence occurs exactly at the transition frequency we show below that it is not problematic.

%
%

To make a higher precision direct measurement of the transition frequency we have planned a multi-phase experiment.  The first phase will be a low resolution study using the Advanced Light Source (ALS) at Lawrence Berkeley Labs.  The ALS provides tunable VUV light (5~eV-30~eV) with a linewidth ($\Delta$) of 0.175~eV, a photon flux ($\Phi_{p}$) of $\sim$10$^{20}$ photons/cm$^2\cdot$s, and a 170~$\mu$m~$\times$~50~$\mu$m spot size.  Assuming both the light ($\Delta$) and broadened transition ($\Gamma$) linewidths to be Lorentzian, then far from saturation, the excitation rate, $\frac{dN_{e}}{dt}$, is given by $\frac{dN_{e}}{dt} \approx  \left( \frac{\lambda}{2 \pi} \right)^2  \frac{\Gamma_n}{\Gamma + \Delta} \frac{1}{1+4(\frac{\omega_0 - \omega_L}{\Gamma + \Delta})^2} \Phi_{p} N_{g}$ where $\omega_0$ ($\lambda$) is the transition frequency (wavelength), $\omega_L$ is the center frequency  of the ALS beam, and $N_{g}$ is the total number of ground state $^{229}$Th nuclei addressed by the ALS beam. 
To locate the transition the ALS center frequency will be scanned across the interesting range of (7.6~$\pm$~0.5~eV) in steps of 0.05~eV with 200~s of  illumination followed by 100~s of light collection at each point.  
Even in the lowest fluorescence case ($\Gamma_n = 2 \pi \times 100$~$\mu$Hz), we expect a minimum total fluorescence rate of $\sim$15~kHz when the ALS is on resonance.  Assuming the $\alpha$-induced background is the dominant noise source this rate allows for a signal to noise ratio (SNR) $>$ 10:1. In this manner, the entire spectral region can be scanned in a few hours.  Because the transition is both narrow and long-lived, backgrounds from scattered light and backgrounds from induced spurious fluorescence, which is in general relatively short-lived and broadband, are not expected to be problematic.  Once the resonance is found, longer illumination will lead to sufficient fluorescence rates for a VUV spectrometer measurement. We expect that the transition frequency can be determined to $\sim$50~cm$^{-1}$ (0.1~nm) in these experiments.

Radiation trapping will limit the total density of $^{229}$Th to $n_{max} = 1 / \sigma \rm w$, where $\sigma$ is the absorption cross section and w is the excitation beam waist ($n_{max}$~$\sim$~10$^{13}$~$^{229}$Th/cm$^3$ for the ALS measurement). However, radiation trapping can be eliminated by applying a sufficient magnetic field gradient ($d$B$/dx$~$\approx$~100~G/cm) across the crystal.  The concomitant line broadening is acceptable in the initial phases of coarsely determining the transition energy.


After the ALS frequency measurement, we will more precisely measure the transition energy  via laser-induced
fluorescence.  A VUV light generation system can be constructed by coupling the output of a pulsed dye laser
into a H$_2$-Raman cell.  With this system the ALS determined frequency range can be covered with a laser pulse energy of 10~$\mu$J and linewidth of 0.1~cm$^{-1}$~\cite{schomburg_generation_1983},
allowing the entire range to be scanned in 6~h with SNR $>$ 10:1.

Following these measurements, evaluation of the transition as an optical frequency reference can begin. The optimum method of precisely
determining the transition frequency is to probe the nuclei with a VUV frequency comb~\cite{jones_phase_2005}.  Locking the
frequency comb to the nuclear transition allows for direct comparison to standard clocks.   Assuming the transition linewidth is 0.5-10~kHz, a 100~pW on-resonance comb tooth leads to a transition frequency resolution of $3 \times 10^{-17} < \Delta f / f < 1 \times 10^{-15}$ with 1~s of photon collection. Longer integration times improve these limits, but only to the level allowed by temperature stability. With the best temperature stability demonstrated to date~\cite{davidowitz_novel_1995} the resolution is limited to $2 \times 10^{-18}$.  This level is reached in 5~min - 5~d for the linewidth range considered.  As mentioned above, the long-term accuracy is $2 \times 10^{-16}$ as set by current thermometer accuracies.


In summary, we have proposed a novel method for the direct measurement of the energy of the low-lying isomeric state in $^{229}$Th. The analysis presented indicates that it should be possible to retain a narrow transition linewidth even when the $^{229}$Th nucleus is doped into a crystal.  The technique is limited by temperature accuracy and stability which allows for a fractional frequency accuracy of $ \Delta f / f \sim 2 \times 10^{-16}$ in the long-term and a resolution of $ \Delta f / f \sim 2 \times 10^{-18}$ in the short-term.  The technique looks promising for a solid-state optical frequency reference, as well as tests of fundamental constant variation.  The sensitive temperature dependence might also make it a useful thermometer.


\acknowledgements{The ALS is supported by the U.S. DOE under Contract No. $DE-AC02–05CH11231$. This work is supported by UCLRP award No. 09-LR-04-120497-HUDE.}


\begin{thebibliography}{26}
\expandafter\ifx\csname natexlab\endcsname\relax\def\natexlab#1{#1}\fi
\expandafter\ifx\csname bibnamefont\endcsname\relax
  \def\bibnamefont#1{#1}\fi
\expandafter\ifx\csname bibfnamefont\endcsname\relax
  \def\bibfnamefont#1{#1}\fi
\expandafter\ifx\csname citenamefont\endcsname\relax
  \def\citenamefont#1{#1}\fi
\expandafter\ifx\csname url\endcsname\relax
  \def\url#1{\texttt{#1}}\fi
\expandafter\ifx\csname urlprefix\endcsname\relax\def\urlprefix{URL }\fi
\providecommand{\bibinfo}[2]{#2}
\providecommand{\eprint}[2][]{\url{#2}}



\bibitem[{\citenamefont{Takamoto and
  Katori}(2003)}]{takamoto_spectroscopy_2003}
\bibinfo{author}{\bibfnamefont{M.}~\bibnamefont{Takamoto}} \bibnamefont{and}
  \bibinfo{author}{\bibfnamefont{H.}~\bibnamefont{Katori}},
  \bibinfo{journal}{Phys. Rev. Lett.} \textbf{\bibinfo{volume}{91}},
  \bibinfo{pages}{223001} (\bibinfo{year}{2003}).

\bibitem[{\citenamefont{Ido et~al.}(2005)\citenamefont{Ido, Loftus, Boyd,
  Ludlow, Holman, and Ye}}]{ido_precision_2005}
\bibinfo{author}{\bibfnamefont{T.}~\bibnamefont{Ido}} \bibnamefont{\textit{et~al.}},
  \bibinfo{journal}{Phys. Rev. Lett.} \textbf{\bibinfo{volume}{94}},
  \bibinfo{pages}{153001} (\bibinfo{year}{2005}).

\bibitem[{\citenamefont{Barber et~al.}(2006)\citenamefont{Barber, Hoyt, Oates,
  Hollberg, Taichenachev, and Yudin}}]{barber_direct_2006}
\bibinfo{author}{\bibfnamefont{Z.~W.} \bibnamefont{Barber}} \bibnamefont{\textit{et~al.}},
  \bibinfo{journal}{Phys. Rev. Lett.} \textbf{\bibinfo{volume}{96}},
  \bibinfo{pages}{083002} (\bibinfo{year}{2006}).

\bibitem[{\citenamefont{Rosenband et~al.}(2007)\citenamefont{Rosenband,
  Schmidt, Hume, Itano, Fortier, Stalnaker, Kim, Diddams, Koelemeij, Bergquist
  et~al.}}]{rosenband_observation_2007}
\bibinfo{author}{\bibfnamefont{T.}~\bibnamefont{Rosenband}} \bibnamefont{\textit{et~al.}},
\bibinfo{journal}{Phys. Rev. Lett.}
  \textbf{\bibinfo{volume}{98}}, \bibinfo{pages}{220801}
  (\bibinfo{year}{2007}).

\bibitem[{\citenamefont{Ashby et~al.}(2007)\citenamefont{Ashby, Heavner,
  Jefferts, Parker, Radnaev, and Dudin}}]{ashby_testing_2007}
\bibinfo{author}{\bibfnamefont{N.}~\bibnamefont{Ashby}} \bibnamefont{\textit{et~al.}},
  \bibinfo{journal}{Phys. Rev. Lett.} \textbf{\bibinfo{volume}{98}},
  \bibinfo{pages}{070802} (\bibinfo{year}{2007}).

\bibitem[{\citenamefont{Rosenband et~al.}(2008)\citenamefont{Rosenband, Hume,
  Schmidt, Chou, Brusch, Lorini, Oskay, Drullinger, Fortier, Stalnaker
  et~al.}}]{rosenband_frequency_2008}
\bibinfo{author}{\bibfnamefont{T.}~\bibnamefont{Rosenband}}  \bibnamefont{\textit{et~al.}},
\bibinfo{journal}{Science}
  \textbf{\bibinfo{volume}{319}}, \bibinfo{pages}{1808} (\bibinfo{year}{2008}).

%
%
%
%
%
%
%
%

\bibitem[{\citenamefont{Tkalya}(2000)}]{tkalya_decay_2000}
\bibinfo{author}{\bibfnamefont{E.~V.}~\bibnamefont{Tkalya}},
\bibinfo{author}{\bibfnamefont{A.~N.}~\bibnamefont{Zherikhin}},
\bibinfo{author}{\bibfnamefont{V.~I.}~\bibnamefont{Zhudov}},
  \bibinfo{journal}{Phys.~Rev.~C} \textbf{\bibinfo{volume}{61}},
  \bibinfo{pages}{064308} (\bibinfo{year}{2000}).

\bibitem[{\citenamefont{Ruchowska}(2006)}]{ruchowska_nuclear_2006}
\bibinfo{author}{\bibfnamefont{E.} \bibnamefont{Ruchowska}} \textit{et al.}
  \bibinfo{journal}{Phys. Rev. C} \textbf{\bibinfo{volume}{73}},
  \bibinfo{pages}{044326} (\bibinfo{year}{200}).

\bibitem[{\citenamefont{Kroger and Reich}(1976)}]{kroger_features_1976}
\bibinfo{author}{\bibfnamefont{L.~A.} \bibnamefont{Kroger}} \bibnamefont{and}
  \bibinfo{author}{\bibfnamefont{C.~W.} \bibnamefont{Reich}},
  \bibinfo{journal}{Nucl. Phys. A} \textbf{\bibinfo{volume}{259}},
  \bibinfo{pages}{29} (\bibinfo{year}{1976}).

\bibitem[{\citenamefont{Helmer and Reich}(1994)}]{helmer_excited_1994}
\bibinfo{author}{\bibfnamefont{R.~G.} \bibnamefont{Helmer}} \bibnamefont{and}
  \bibinfo{author}{\bibfnamefont{C.~W.} \bibnamefont{Reich}},
  \bibinfo{journal}{Phys. Rev. C} \textbf{\bibinfo{volume}{49}},
  \bibinfo{pages}{1845} (\bibinfo{year}{1994}).

\bibitem[{\citenamefont{Irwin and Kim}(1997)}]{irwin_observation_1997}
\bibinfo{author}{\bibfnamefont{G.~M.} \bibnamefont{Irwin}} \bibnamefont{and}
  \bibinfo{author}{\bibfnamefont{K.~H.} \bibnamefont{Kim}},
  \bibinfo{journal}{Phys. Rev. Lett.} \textbf{\bibinfo{volume}{79}},
  \bibinfo{pages}{990} (\bibinfo{year}{1997}).

\bibitem[{\citenamefont{Richardson et~al.}(1998)\citenamefont{Richardson,
  Benton, Evans, Griffith, and Tungate}}]{richardson_ultraviolet_1998}
\bibinfo{author}{\bibfnamefont{D.~S.} \bibnamefont{Richardson}} \bibnamefont{\textit{et~al.}},
  \bibinfo{journal}{Phys. Rev. Lett.} \textbf{\bibinfo{volume}{80}},
  \bibinfo{pages}{3206} (\bibinfo{year}{1998}).

\bibitem[{\citenamefont{Beck et~al.}(2007)\citenamefont{Beck, Becker,
  Beiersdorfer, Brown, Moody, Wilhelmy, Porter, Kilbourne, and
  Kelley}}]{beck_energy_2007}
\bibinfo{author}{\bibfnamefont{B.~R.} \bibnamefont{Beck}} \bibnamefont{\textit{et~al.}},
\bibinfo{journal}{Phys. Rev. Lett.}
  \textbf{\bibinfo{volume}{98}}, \bibinfo{pages}{142501}
  (\bibinfo{year}{2007}).

\bibitem[{\citenamefont{Hudson et~al.}(2008)\citenamefont{Hudson, Vutha,
  Lamoreaux, and DeMille}}]{hudson_investigation_2008}
\bibinfo{author}{\bibfnamefont{E.~R.} \bibnamefont{Hudson}} \bibnamefont{\textit{et~al.}},
  \bibinfo{journal}{\textit{International Conference on Atomic Physics}}, (\bibinfo{year}{2008}),
  \bibinfo{volume}{Poster {MO28}}.

\bibitem[{\citenamefont{Peik}(2009)}]{peik_prospects_2009}
\bibinfo{author}{\bibfnamefont{E.}~\bibnamefont{Peik}}
\bibnamefont{\textit{et~al.}},
  \bibinfo{journal}{arXiv:0812.3548v2} (\bibinfo{year}{2009}).

\bibitem[{\citenamefont{Peik and Tamm}(2003)}]{peik_nuclear_2003}
\bibinfo{author}{\bibfnamefont{E.}~\bibnamefont{Peik}} \bibnamefont{and}
  \bibinfo{author}{\bibfnamefont{C.}~\bibnamefont{Tamm}},
  \bibinfo{journal}{Europhys. Lett.}
  \textbf{\bibinfo{volume}{61}}, \bibinfo{pages}{181} (\bibinfo{year}{2003}).



\bibitem[{\citenamefont{Steele et~al.}(2008)\citenamefont{Steele, Campbell, Churchill, DePalatis, Naylor, Kuzmich, and Chapman}}]{steele_laser_2008}
\bibinfo{author}{\bibfnamefont{A.} \bibnamefont{Steele}} \bibnamefont{\textit{et~al.}},
  \bibinfo{journal}{\textit{Division of Atomic, Molecular, and Optical Physics}}, (\bibinfo{year}{2008}),
  \bibinfo{volume}{R1.00095}.


\bibitem[{\citenamefont{Dicke}(1953)}]{dicke_effect_1953}
\bibinfo{author}{\bibfnamefont{R.~H.}~\bibnamefont{Dicke}},
  \bibinfo{journal}{Phys. Rev. B} \textbf{\bibinfo{volume}{89}},
  \bibinfo{pages}{472} (\bibinfo{year}{1953}).


\bibitem[{\citenamefont{Casten}(1990)}]{casten_nuclear_1990}
\bibinfo{author}{\bibfnamefont{R.}~\bibnamefont{Casten}},
  \emph{\bibinfo{title}{Nuclear structure from a simple perspective}}
  (\bibinfo{publisher}{Oxford University Press}, \bibinfo{address}{New York},
  \bibinfo{year}{1990}).


\bibitem[{\citenamefont{Gibb}(1980)}]{gibb_principles_1980}
\bibinfo{author}{\bibfnamefont{T.~C.} \bibnamefont{Gibb}},
  \emph{\bibinfo{title}{Principles of M\"ossbauer spectroscopy, Studies in
  chemical physics}} (\bibinfo{publisher}{Chapman and Hall},
  \bibinfo{address}{London}, \bibinfo{year}{1980}).


\bibitem[{\citenamefont{schroeer}(1979)}]{schroeer_isomer_1979}
\bibinfo{author}{\bibfnamefont{D.}~\bibnamefont{Schroeer}}, \bibinfo{author}{\bibfnamefont{Ch.~S.}~\bibnamefont{Kuo}} \bibnamefont{and}
  \bibinfo{author}{\bibfnamefont{R.~J.}~\bibnamefont{Lambe}},
  \bibinfo{journal}{Phys. Status Solidi B} \textbf{\bibinfo{volume}{92}},
  \bibinfo{pages}{565} (\bibinfo{year}{1979}).

\bibitem[{\citenamefont{rohde}(1983)}]{rohde_temperature_1983}
\bibinfo{author}{\bibfnamefont{M.}~\bibnamefont{Rohde}},  \bibnamefont{and}
  \bibinfo{author}{\bibfnamefont{D.}~\bibnamefont{Salomon}},
  \bibinfo{journal}{Hyperfine Interact.} \textbf{\bibinfo{volume}{15}},
  \bibinfo{pages}{257} (\bibinfo{year}{1983}).


\bibitem[{\citenamefont{davidowitz}(1994)}]{davidowitz_novel_1995}
\bibinfo{author}{\bibfnamefont{H.}~\bibnamefont{Davidowitz}},
  \bibinfo{journal}{Rev. Sci. Instrum.} \textbf{\bibinfo{volume}{67}},
  \bibinfo{pages}{1} (\bibinfo{year}{1994}).


\bibitem[{\citenamefont{Dykhe and Tkalya}(1998)}]{dykhne_matrix_1997}
\bibinfo{author}{\bibfnamefont{A.}~\bibnamefont{Dykhne}} \bibnamefont{and}
  \bibinfo{author}{\bibfnamefont{E.}~\bibnamefont{Tkalya}},
  \bibinfo{journal}{JETP Lett.} \textbf{\bibinfo{volume}{67}},
  \bibinfo{pages}{251} (\bibinfo{year}{1998}).

\bibitem[{\citenamefont{Anisimov}(2007)}]{anisimov_concept_2007}
\bibinfo{author}{\bibfnamefont{P.}~\bibnamefont{Anisimov}},
\bibinfo{author}{\bibfnamefont{Y.}~\bibnamefont{Rostovtsev}},
\bibinfo{author}{\bibfnamefont{O.}~\bibnamefont{Kocharovskaya}},
  \bibinfo{journal}{Phys. Rev. B} \textbf{\bibinfo{volume}{76}},
  \bibinfo{pages}{094422} (\bibinfo{year}{2007}).


\bibitem[{\citenamefont{Gerstenkorn}(1974)}]{gerstenkorn_hyperfine_1974}
\bibinfo{author}{\bibfnamefont{S.}~\bibnamefont{Gerstenkorn}},
  \bibinfo{journal}{J. Phys. (Paris)} \textbf{\bibinfo{volume}{35}},
  \bibinfo{pages}{483} (\bibinfo{year}{1974}).





\bibitem[{\citenamefont{Ramakrishna}(1966)}]{ramakrishna_electric_1966}
\bibinfo{author}{\bibfnamefont{J.}~\bibnamefont{Ramakrishna}},
  \bibinfo{journal}{Philos. Mag.} \textbf{\bibinfo{volume}{13}},
  \bibinfo{pages}{515} (\bibinfo{year}{1966}).



\bibitem[{\citenamefont{Roesch}(1975)}]{roesch_electric_1975}
\bibinfo{author}{\bibfnamefont{L.~Ph.}~\bibnamefont{Roesch}},
\bibinfo{author}{\bibfnamefont{R.}~\bibnamefont{Kulessa}},
\bibnamefont{and}
\bibinfo{author}{\bibfnamefont{F.}~\bibnamefont{Horber}},
  \bibinfo{journal}{Phys. Status Solidi B } \textbf{\bibinfo{volume}{71}},
  \bibinfo{pages}{389} (\bibinfo{year}{1975}).

\bibitem[{\citenamefont{Slack}(1961)}]{slack_thermal_1961}
\bibinfo{author}{\bibfnamefont{Glen A.}~\bibnamefont{Slack}},
  \bibinfo{journal}{Phys. Rev.} \textbf{\bibinfo{volume}{122}},
  \bibinfo{pages}{1451} (\bibinfo{year}{1961}).


\bibitem[{\citenamefont{Knop}(1975)}]{knop_arrangements_1975}
\bibinfo{author}{\bibfnamefont{O.}~\bibnamefont{Knop}},
\bibinfo{author}{\bibfnamefont{E.~M.}~\bibnamefont{Palmer}}, \bibnamefont{and}
\bibinfo{author}{\bibfnamefont{R.~W.}~\bibnamefont{Robinson}},
  \bibinfo{journal}{Acta Cryst.} \textbf{\bibinfo{volume}{31}},
  \bibinfo{pages}{19} (\bibinfo{year}{1975}).

\bibitem[{not({\natexlab{a}})}]{ITS90}
\bibinfo{title}{H. Preston-Thomas, Metro. \textbf{27}, 3 (1990) and on-line supplement}.


\bibitem[{\citenamefont{Flambaum and Kozlov}(2007)}]{flambaum_enhanced_2007}
\bibinfo{author}{\bibfnamefont{V.~V.} \bibnamefont{Flambaum}} \bibnamefont{and}
  \bibinfo{author}{\bibfnamefont{M.~G.} \bibnamefont{Kozlov}},
  \bibinfo{journal}{Phys. Rev. Lett.} \textbf{\bibinfo{volume}{99}},
  \bibinfo{pages}{150801} (\bibinfo{year}{2007}).


%



\bibitem[{\citenamefont{tao He and zhou Ren}(2007)}]{he_enhanced_2007}
\bibinfo{author}{\bibfnamefont{X.}~\bibnamefont{tao He}} \bibnamefont{and}
  \bibinfo{author}{\bibfnamefont{Z.}~\bibnamefont{zhou Ren}},
  \bibinfo{journal}{J. Phys. G: Nucl. Part. Phys.}
  \textbf{\bibinfo{volume}{34}}, \bibinfo{pages}{1611} (\bibinfo{year}{2007}).

\bibitem[{\citenamefont{Flambaum}(2006)}]{flambaum_enhanced_2006}
\bibinfo{author}{\bibfnamefont{V.~V.} \bibnamefont{Flambaum}}
  \bibinfo{journal}{Phys. Rev. Lett.} \textbf{\bibinfo{volume}{97}},
  \bibinfo{pages}{092502} (\bibinfo{year}{2006}).

   \bibitem[{\citenamefont{Flambaum}(2009)}]{flambaum_enhanced_2009}
\bibinfo{author}{\bibfnamefont{V.~V.} \bibnamefont{Flambaum}} \bibnamefont{and}
  \bibinfo{author}{\bibfnamefont{R.~B.} \bibnamefont{Wiringa}},
  \bibinfo{journal}{Phys. Rev. C} \textbf{\bibinfo{volume}{79}},
  \bibinfo{pages}{034302} (\bibinfo{year}{2009}).

\bibitem[{\citenamefont{Flambaum}(2008)}]{flambaum_variation_2008}
\bibinfo{author}{\bibfnamefont{V.~V.} \bibnamefont{Flambaum}},
  \bibinfo{journal}{ Eur. Phys. J. Special Topics} \textbf{\bibinfo{volume}{163}},
  \bibinfo{pages}{159} (\bibinfo{year}{2008}).












\bibitem[{\citenamefont{Burkhalter et~al.}(2001)\citenamefont{Burkhalter,
  Dohnke, and Hulliger}}]{Burkhalter2001}
\bibinfo{author}{\bibfnamefont{R.}~\bibnamefont{Burkhalter}},
  \bibinfo{author}{\bibfnamefont{I.}~\bibnamefont{Dohnke}}, \bibnamefont{and}
  \bibinfo{author}{\bibfnamefont{J.}~\bibnamefont{Hulliger}},
  \bibinfo{journal}{Progress in Crystal Growth and Characterization of
  Materials} \textbf{\bibinfo{volume}{42}}, \bibinfo{pages}{1}
  (\bibinfo{year}{2001}).

\bibitem[{\citenamefont{Samtleben and Hulliger}(2005)}]{Samtleben2005}
\bibinfo{author}{\bibfnamefont{T.~A}~\bibnamefont{Samtleben}} \bibnamefont{and}
  \bibinfo{author}{\bibfnamefont{J.}~\bibnamefont{Hulliger}},
  \bibinfo{journal}{Optics and Lasers in Engineering}
  \textbf{\bibinfo{volume}{43}}, \bibinfo{pages}{251} (\bibinfo{year}{2005}).

\bibitem[{\citenamefont{Sabatini et~al.}(1975)\citenamefont{Sabatini, Salwin,
  and McClure}}]{Sabatini1975}
\bibinfo{author}{\bibfnamefont{J.~F.}~\bibnamefont{Sabatini}},
  \bibinfo{author}{\bibfnamefont{A.~E.}~\bibnamefont{Salwin}}, \bibnamefont{and}
  \bibinfo{author}{\bibfnamefont{D.~S.}~\bibnamefont{McClure}},
  \bibinfo{journal}{Phys. Rev. B} \textbf{\bibinfo{volume}{11}},
  \bibinfo{pages}{3832} (\bibinfo{year}{1975}).
%
%

\bibitem[{\citenamefont{Hehlen}(2009)}]{Hehlen2009}
\bibinfo{author}{\bibfnamefont{M.~P.}~\bibnamefont{Hehlen}}, in
  \emph{\bibinfo{booktitle}{{Optical Refrigeration. Science and Applications of
  Laser Cooling of Solids}}}
  (\bibinfo{publisher}{Wiley, Weinheim}, \bibinfo{year}{2009}).


\bibitem[{\citenamefont{Jackson}(2008)}]{jackson_private_2008}
\bibinfo{author}{\bibfnamefont{R.}~\bibnamefont{Jackson} \textit{et al.}},
  \bibinfo{journal}{J. Phys. Condens. Matter }
\textbf{\bibinfo{volume}{21}},
  \bibinfo{pages}{3245403}
 (\bibinfo{year}{2009}).















\bibitem[{\citenamefont{Schomburg et~al.}(1983)\citenamefont{Schomburg,
  Dobele, and Ruckle}}]{schomburg_generation_1983}
\bibinfo{author}{\bibfnamefont{H.}~\bibnamefont{Schomburg}},
  \bibinfo{author}{\bibfnamefont{H.~F.} \bibnamefont{D\"obele}},
  \bibnamefont{and} \bibinfo{author}{\bibfnamefont{B.}~\bibnamefont{R\"uckle}},
  \bibinfo{journal}{Appl Phys B: Lasers Opt} \textbf{\bibinfo{volume}{30}},
  \bibinfo{pages}{131} (\bibinfo{year}{1983}).


\bibitem[{\citenamefont{Jones et~al.}(2005)\citenamefont{Jones, Moll, Thorpe, and Ye}}]{jones_phase_2005}
\bibinfo{author}{\bibfnamefont{R.~J.}~\bibnamefont{Jones}} \bibnamefont{\textit{et~al.}},
  \bibinfo{journal}{Phys. Rev. Lett.}
  \textbf{\bibinfo{volume}{94}}, \bibinfo{pages}{193201} (\bibinfo{year}{2005}).




\end{thebibliography}
\end{document}